\documentclass[10pt]{article}

\usepackage{amssymb}
\usepackage{amsmath}
\usepackage{graphicx}

\begin{document}

\begin{center}
\Large{\bf TUNNELING AS A CLASSICAL ESCAPE RATE INDUCED BY THE VACUUM
ZERO-POINT RADIATION}\\[0.3cm]
\end{center}

\vspace{0.5cm}

\begin{center}
A. J. FARIA, H. M. FRAN\c{C}A\footnote{e-mail: hfranca@if.usp.br},
and R. C. SPONCHIADO \\[0.3cm]
\end{center}

\begin{center}
{\em Instituto de F\'{i}sica, Universidade de S\~{a}o Paulo \\
C.P. 66318, 05315-970 S\~{a}o Paulo, SP, Brazil}
\end{center}

\vspace{0.5cm}

\begin{abstract}
We make a brief review of the Kramers escape rate theory for the
probabilistic motion of a particle in a potential well $U(x)$, and
under the influence of classical fluctuation forces. The Kramers
theory is extended in order to take into
account the action of the thermal and zero-point random
electromagnetic fields on a charged particle. The result is
physically relevant because we get a non null escape rate over the
potential barrier at low temperatures ($T \rightarrow 0$). It is
found that, even if the mean energy is much smaller than the
barrier height, the classical particle can escape from the
potential well due to the action of the zero-point fluctuating
fields. These stochastic effects can be used to give a
{\em classical} interpretation to some {\em quantum tunneling}
phenomena. Relevant experimental data are used to illustrate the
theoretical results.
\end{abstract}


\vspace{2cm}

\section{Introduction}

One of the most useful contributions to our understanding of the
stochastic processes theory is the study of escape rates over a
potential barrier. The theoretical approach, first proposed by
Kramers \cite{Kramers}, has many applications in chemistry
kinetics, diffusion in solids, nucleation \cite{Hanggi}, and other phenomena
\cite{Santos}. The essential structure of the escape process is
that the bounded particle is under the action of three types of
forces: a deterministic nonlinear force with at least one
metastable region, a fluctuating force whose action is capable of
pushing the particle out of the metastable region, and a
dissipative force which inevitably accompany the fluctuations.

In this work we describe the escape rates of a particular model: a
classical charged particle moving in the metastable potential shown in
the Figure 1, and under the influence of the fluctuating
electromagnetic radiation forces commonly used in Stochastic
Electrodynamics (SED)\cite{Marshall,Boyer1}. The fluctuating fields postulated
in SED are classical random fields, with zero mean but nonzero higher
moments. The spectral distribution of this radiation can be expressed
as a sum of two terms
\begin{eqnarray}
\rho(\omega,T) & = & \frac{\omega^{2}}{\pi^{2}c^{3}} \left[
\frac{\hbar \omega}{2}+ \frac{\hbar \omega}{e^{\hbar
\omega/k_B T}-1}\right]
\label{1}
\end{eqnarray}
The first term is the zero-point radiation contribution to the
spectral distribution. It is independent of the temperature and is
Lorentz invariant. The second term in (\ref{1}) is the blackbody
radiation spectral distribution, responsible for the temperature
effects on the system.

\begin{figure}\label{fig1}
\centerline{\includegraphics[width=80mm]{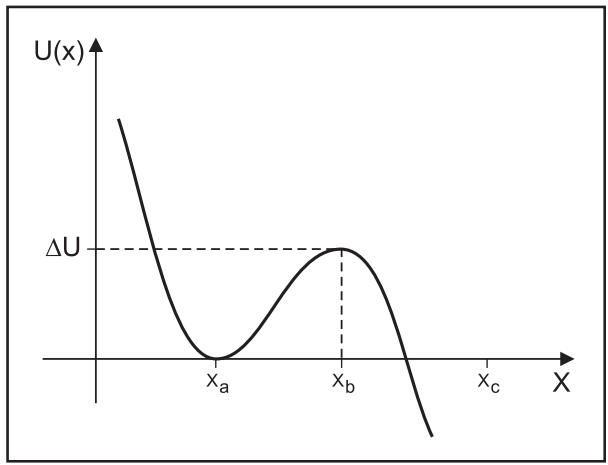}}
\caption{Metastable potential with a barrier height $\Delta U=
U(x_b)-U(x_a)$, a local minimum $x_{a}$ and a local maximum
$x_{b}$ (top of the barrier).}
\end{figure}

\section{Properties of the harmonic oscillator motion under the action of a thermal and zero-point radiation}

The zero-point radiation (first term in (\ref{1})) has a mean energy $ \hbar \omega/2$ associated with each mode of the
electromagnetic fields, and is responsible for the most important features of SED. With this zero-point radiation
postulated, several phenomena associated with the quantum behavior of the microscopic world can be explained on
classical grounds. Many examples can be found in the reviews
\cite{Boyer1,Boyer2,Pena2,Milonni1}.

The potential $U(x)$ will be approximated by a harmonic oscillator in
the region of the potential well $(x \approx x_a)$ so that (see Figure
1)
\begin{equation} \label{Ua}
U(x) \simeq U(x_a) + \frac{1}{2} m\omega_a^2 (x - x_a)^2 ,
\end{equation}
where $\omega_a$ is the natural frequency of the oscillator.

The dynamical behavior of a harmonically bounded charged particle has
been extensively studied in the context of
classical SED. It is found that the zero-point radiation maintains the
stability of this system. We shall use the statistical properties of the
harmonic oscillator in order to understand, classically, the escape
rate at very low temperatures. We give below a brief review of the
harmonic motion under the action of the random electric fields
characteristic of SED.

The nonrelativistic motion of the charged particle (charge $e$
and mass $m$) near the bottom of the potential well (see Figure 1) is
governed by the equation
\begin{equation}\label{20}
m \ddot{\xi} = -m \omega_a^2 \xi + \frac{2e^2}{3c^3}
\stackrel{\ldots}{\xi} + e E_x(t) ,
\end{equation}
where $\xi = x - x_a$, and $E_x(t)$ is the $x$ component of the random
electric field. The term proportional to $\stackrel{\ldots}{\xi}$ is
the radiation reaction force. The electric field is such that
$\left\langle E_x(t) \right\rangle = 0$ and
\begin{equation} \label{21}
\left\langle E_{x}(t) E_{x}(0) \right\rangle = \frac{4 \pi}{3}
\int^{\infty}_{0} d\omega \rho(\omega,T) \cos(\omega t) ,
\end{equation}
where the spectral distribution $\rho(\omega,T)$ was introduced in
the equation (\ref{1}). The radiation reaction force can be
approximated by \cite{jackson}
\begin{equation}\label{21B}
\frac{2e^2}{3c^3} \stackrel{\ldots}{\xi} \simeq -m \gamma \dot{\xi} ,
\end{equation}
where $\gamma = 2e^2 \omega_a^2 / 3mc^3$. Moreover it is verified that
$\gamma \ll \omega_{a}$. According to these approximations one can
show that the average energy $\langle \epsilon \rangle$ of the
oscillating charge is such that
\begin{equation}\label{D}
\langle \epsilon \rangle =
\frac{1}{2}m \langle \dot{\xi}^2 \rangle +
\frac{1}{2}m \omega_a^2 \langle \xi^{2} \rangle =
\frac{\hbar\omega_a}{2} \coth \left( \frac{\hbar\omega_a}{2k_B T} \right)
\equiv D(T) ,
\end{equation}
where we have introduced the function $D(T)$ in order to
simplify our notation. Notice that the average energy depends on the
temperature and on the oscillatory frequency $\omega_a$.

The result (\ref{D}) is well known \cite{Boyer1}. The
average energy $\langle \epsilon \rangle$ becomes equal to
$k_B T$ in the high temperature limit ($k_B T \gg \hbar \omega_a$), and is
non zero when $T = 0$. Actually $D(T) \rightarrow \hbar \omega_a/2$
as $T \rightarrow 0$. Notice that $D(T)$ depends on $\hbar$. We can
show that the Planck constant comes from the intensity of the
zero-point field $E_x(t)$ that appears in (\ref{20}). We recall that
$\hbar \omega_a/2$ is the value of the ground state energy of
the harmonic oscillator in quantum mechanics.
This result, obtained within the realm of SED,
differs from the usual null result of ordinary classical physics
because the zero-point fluctuations are taken into account.

It is quite natural to use the average energy (\ref{D}) in
the calculation of the Kramers escape rate of a potential well. We shall
see that the consequence of the new form of the average energy is
a non-vanishing escape rate even if $T \rightarrow 0$. Most
authors do not mention the classical zero-point fluctuations and use the
quantum mechanical formalism to interpret the non null escape
rate as a tunneling through the classical forbidden region of
the barrier. We shall see that the zero-point fluctuations
allow the escape over the potential barrier even if the mean
energy of the particle inside the barrier is {\em much less} than
$\Delta U=U(x_b)-U(x_a)$. In the classical mechanics context the
escape would be impossible without the action of the zero-point
fluctuations. For simplicity we shall take $U(x_a)=0$ in what
follows.

\section{The escape rate over the potential well}

We shall use the approach of Chandrasekhar \cite{Chandrasekhar},
based on the Kramers theory.
The physical system considered by Chandrasekhar is a particle moving
under the influence of a fluctuating force, and a potential $U(x)$
that has a metastable region (see Figure 1). The motion of the
particle is governed by a Langevin type equation
\begin{equation} \label{A}
m \ddot{x}= -m \gamma \dot{x} - U'(x) + F(t) ,
\end{equation}
where $-m \gamma \dot{x}$ is the dissipative force and $F(t)$ is
the fluctuating force which is characterized by the average
$\langle F(t) \rangle = 0 \nonumber $. The average energy of the
particle within the potential well, that is $x < x_b$, is assumed to be
given by
\begin{equation}\label{19}
\langle \frac{m}{2} \dot{x}^{2} + U(x) \rangle = k_B T ,
\end{equation}
in the high temperature limit.

It is possible to show that the Langevin equation (\ref{A}) leads to
a phase space Fokker-Planck equation given by \cite{Chandrasekhar}
\begin{equation} \label{FP}
\frac{\partial W}{\partial t} +
\frac{p}{m} \frac{\partial W}{\partial x} -
U'(x) \frac{\partial W}{\partial p} =
\gamma W + \gamma p \frac{\partial W}{\partial p} +
m \gamma k_B T \frac{\partial^2 W}{\partial p^2} ,
\end{equation}
where $W=W(x,p,t)$ is the probability distribution in
phase space. Notice that the left hand side of the above expression is
equivalent to the Liouville equation. The right hand side appears as a
consequence of the fluctuating and dissipation forces. For low
temperatures, the Fokker-Planck equation (\ref{FP}) is not
valid. As we have mentioned in previous section (see the equation
(\ref{D})), the factor $k_B T$ in the last term of (\ref{FP}) must be
replaced by $D(T) = \frac{\hbar\omega_{a}}{2} \coth\left(
\frac{\hbar\omega_{a}}{2k_B T} \right)$. Therefore, we shall consider the
following equation
\begin{equation} \label{FPD}
\frac{\partial W}{\partial t} +
\frac{p}{m} \frac{\partial W}{\partial x} -
U'(x) \frac{\partial W}{\partial p} =
\gamma W + \gamma p \frac{\partial W}{\partial p} +
m \gamma D(T) \frac{\partial^2 W}{\partial p^2} .
\end{equation}
We shall see that the equation (\ref{FPD}) will allow us to give an
accurate description of the escape rate at low temperatures.

In the Kramers theory, two quantities are essential to calculate the
escape rate. One is the probability $P(t)$ of finding the particle
inside the potential well. This probability can be obtained from the
phase space distribution, namely
\begin{equation}\label{P}
P(t) = \int_{-\infty}^{\infty}dp \int_{-\infty}^{x_b}dx W(x,p,t) .
\end{equation}
The other important quantity is the diffusion current, $j(x_b)$,
across the top of the potential barrier. The diffusion current in an
arbitrary position $x$ is defined by
\begin{equation}\label{def1}
j(x,t) \equiv \int_{-\infty}^{\infty} dp \frac{p}{m} W(x,p,t) .
\end{equation}

Using (\ref{FPD}), (\ref{P}) and (\ref{def1}) one can show that
\begin{equation}\label{P2}
\frac{\partial P(t)}{\partial t} = 
- \int_{-\infty}^{\infty}dp \frac{p}{m} W(x_b,p,t) = -j(x_b,t) .
\end{equation}

The escape rate $\kappa$, regarded as the decay factor of the
probability $P(t)$, can be defined by the equation
\begin{equation}\label{def0}
\frac{\partial P(t)}{\partial t} = - \kappa P(t) .
\end{equation}
The solution of the above equation is
\begin{equation}\label{P0}
P(t)=P_0 e^{-\kappa t} ,
\end{equation}
where $P_0$ is a constant that will be calculated later. On the other
hand, consistently with the equations (\ref{P2}) and (\ref{def0}), one
can define the escape rate as
\begin{equation}\label{def2}
\kappa \equiv \frac{j(x_b,t)}{P(t)} .
\end{equation}
Therefore, according to the above theory we have 
\begin{equation}\label{WQ}
W(x,p,t) = Q(x,p) e^{-\kappa t} ,
\end{equation}
where $Q(x,p)$ satisfies the equation
\begin{equation} \label{FPQ}
\frac{p}{m} \frac{\partial Q}{\partial x} -
U'(x) \frac{\partial Q}{\partial p} -
\gamma Q - \gamma p \frac{\partial Q}{\partial p} -
m \gamma D(T) \frac{\partial^2 Q}{\partial p^2} = 0 .
\end{equation}

A physically interesting case could be
\begin{equation}\label{MB}
Q(x,p) \propto \exp \left[ -\frac{p^2/2m + U(x)}{D(T)} \right] ,
\end{equation}
however, this
standard distribution leads to a situation in which there is no
diffusion across the potential barrier at $x_b$.

Under the conditions of our problem, the equilibrium distribution
(\ref{MB}) cannot be valid for all values of $x$. Hence we shall
consider a solution of the equation (\ref{FPQ}) in the following form 
\begin{equation}\label{MB2}
Q(x,p)= C F(x,p) \exp \left[ -\frac{\frac{p^2}{2m} +
\frac{1}{2} m \omega_a^2 (x-x_a)^2}{D(T)} \right] ,
\end{equation}
where $F(x,p)$ is an unknown function that will be determined
below, and $C$ is a normalization constant. Notice that the above
expression is valid for the phase space motion near the bottom of the
potential well (see Figure 1). The new form for $Q(x,p)$, introduced
in (\ref{MB2}), requires a boundary condition on $F(x,p)$ for $ x
\approx x_a$, namely $F(x,p) \simeq 1$. An alternative form for this
boundary condition is \cite{Chandrasekhar}
\begin{equation}\label{BC1}
F(x,p) \rightarrow 1 , \,\,\, {\rm for}\,\,\, x \ll x_b .
\end{equation}

Another  physical hypothesis is necessary. The probabilistic motion
near the top of the barrier $(x \approx x_b)$ is also governed by a
phase space distribution similar to (\ref{MB2}), namely
\begin{equation}\label{MB3}
Q(x,p) = C F(x,p) \exp \left[ -\frac{\frac{p^2}{2m} + U(x_b) -
\frac{1}{2} m \omega_b^2 (x-x_b)^2}{D(T)} \right] ,
\end{equation}
because
\begin{equation}\label{MB3b}
U(x) \simeq U(x_b) - \frac{1}{2} m \omega_b^2 (x - x_b)^2 .
\end{equation}
The use of $D(T)$ in both formulas (\ref{MB2}) and (\ref{MB3}) is
justified because we are assuming that the particle stays a {\em long
time} in the potential well (see section 2), and crosses the top of the
barrier very quickly ($\omega_b \gg \omega_a$).

Since only a few particles can escape over the potential barrier,
another boundary condition must be imposed on $F(x,p)$, that is
\begin{equation}\label{BC2}
F(x,p) \rightarrow 0 , \,\,\, {\rm for}\,\,\, x \gg x_b .
\end{equation}
Notice that the boundary conditions (\ref{BC1}) and (\ref{BC2}) are simple
hypothesis that can be justified on physical grounds.

Substituting (\ref{MB3}) into (\ref{FPQ}), we obtain the following
differential equation for $F(x,p)$,
\begin{equation}\label{FP2}
\frac{p}{m} \frac{\partial F}{\partial x} +
m \omega_b^2 (x-x_b) \frac{\partial F}{\partial p} =
- \gamma p \frac{\partial F}{\partial p} +
m \gamma D(T) \frac{\partial^2 F}{\partial p^2} .
\end{equation}

Following Chandrasekhar we assume that $F(x,p) = F \left( p - \alpha
m (x-x_b) \right) \equiv F(y)$, where $\alpha$ will be obtained
below. With the introduction of the variable $y$, we obtain the more
simple differential equation
\begin{equation}\label{FP4}
-(\alpha-\gamma) y \frac{d F}{d y} = m \gamma D(T) \frac{d^2 F}{d y^2} ,
\end{equation}
provided that the constant $\alpha$ is such that
\begin{equation}\label{a}
\alpha = \frac{\omega_b^2}{\alpha-\gamma} .
\end{equation}
This equation for the constant $\alpha$ has the solutions
\begin{equation}\label{root}
\alpha = \frac{\gamma}{2} \pm \sqrt{\frac{\gamma^2}{4}+\omega_b^2} .
\end{equation}

The single variable differential equation (\ref{FP4}) can be integrated giving
\begin{equation}\label{F1}
F = F_0 \int^y_{-\infty} \exp \left[ -\frac{ (\alpha-\gamma) y^{\prime
\, 2}}{2m \gamma D(T)} \right] dy^{\prime} ,
\end{equation}
where $F_0$ is a constant. One can see that
only the positive root in (\ref{root}) leads to $\alpha-\gamma$
positive, so that $F(y)$ naturally obeys the boundary conditions
(\ref{BC1}) and (\ref{BC2}). Therefore, one can show that
\begin{equation}\label{F2}
F(x,p) = \sqrt{\frac{\alpha-\gamma}{2 \pi m \gamma D(T)}}
\int^y_{-\infty} \exp \left[ -\frac{(\alpha-\gamma)y^{\prime \, 2}}{2m
\gamma D(T)}\right] dy^{\prime} .
\end{equation}

Combining (\ref{MB3}) and (\ref{F2}) we get for $Q(x,p)$ the result
\begin{eqnarray}\label{MB4}
& & Q(x,p) = C \sqrt{\frac{\alpha-\gamma}{2 \pi m \gamma D(T)}}
\times \\ & & \times
\exp \left[ -\frac{\frac{p^2}{2m} + U(x_b) - \frac{1}{2} m
\omega_b^2 (x-x_b)^2}{D(T)} \right]
\int^y_{-\infty} \exp \left[ -\frac{(\alpha-\gamma)y^{\prime \, 2}}{2m
\gamma D(T)} \right]dy^{\prime} . \nonumber
\end{eqnarray}
The equation (\ref{MB4}) is valid only in the neighborhood of
$x_b$. Inside the potential well ($x \approx x_a$) the approximate
solution is (see section 2)
\begin{equation}\label{MB5}
Q(x,p) = C \exp \left[ - \frac{\frac{p^2}{2m} +
\frac{1}{2} m \omega_a^2 (x-x_a)^2}{D(T)} \right] .
\end{equation}

Using the expression (\ref{MB5}), and considering the equations
(\ref{P}), (\ref{P0}) and (\ref{WQ}), we obtain for the constant $P_0$
\begin{eqnarray}\label{MB6}
P_0 & = & \int_{-\infty}^{\infty}dp \int_{-\infty}^{\infty}dx
C \exp \left[ - \frac{\frac{p^2}{2m} +
\frac{1}{2} m \omega_a^2 (x-x_a)^2}{D(T)} \right]
= \nonumber \\
& = & C \frac{2 \pi D(T)}{\omega_a} .
\end{eqnarray}

The diffusion current across the top of the barrier is (see (\ref{def1})
and (\ref{WQ}))
\begin{equation}\label{j}
j(x_b) = \int_{-\infty}^{\infty} dp \frac{p}{m} Q(x=x_b,p) ,
\end{equation}
where $Q(x=x_b,p)$ is given by our expression (\ref{MB4}). From (\ref{j})
we get
\begin{eqnarray}\label{MB8}
j(x_b) & = & C \sqrt{\frac{a-\gamma}{2 \pi m \gamma D(T)}} 
\exp \left[ -\frac{U(x_b)}{D(T)} \right]
\times \\ & \times & 
\int_{-\infty}^{\infty}dp \frac{p}{m}
\exp \left[ -\frac{p^2}{2mD(T)} \right]
\int^{p}_{-\infty}dy^{\prime}
\exp \left[ -\frac{(\alpha-\gamma)y^{\prime \, 2}}{2m \gamma D(T)} \right].
\nonumber
\end{eqnarray}
It is straightforward to show that
\begin{equation}\label{MB10}
j(x_b) = C \sqrt{\frac{\alpha-\gamma}{\alpha}}D(T)
\exp \left[ -\frac{U(x_b)}{D(T)} \right] .
\end{equation}

The escape rate $\kappa$, defined in (\ref{def2}), becomes
\begin{equation}\label{lowk}
\kappa = \frac{j(x_b)}{P_0} =
\frac{\omega_a}{2\pi} \sqrt{\frac{\alpha-\gamma}{\alpha}}
\exp \left[ -\frac{U(x_b)}{D(T)} \right] ,
\end{equation}
independently of the normalization constant $C$. Notice that the
exponential factor $e^{-\kappa t}$, present in $j(x_b,t)$ and $P(t)$,
cancels leading to the result (\ref{lowk}).

From the expression (\ref{root}) for the positive root, it is possible
to show that 
\begin{equation}\label{lowk2}
\kappa = \frac{\omega_a}{2\pi\omega_b}
\left(\sqrt{\frac{\gamma^2}{4}+\omega_b^2}-\frac{\gamma}{2}\right)
\exp \left[ -\frac{\Delta U}{D(T)} \right] .
\end{equation}
Notice that in the low friction limit $\gamma \rightarrow 0$, this
expression is the simple formula  $\kappa \simeq (\omega_a/2
\pi) \exp{(-\Delta U/D)}$.
It is very important to remark that, in this equation, the escape rate
depends on the potential height $\Delta U$, and on the parameters
characterizing the particle motion {\em inside} the barrier, namely,
the frequency $\omega_a$ and the average energy $\langle \epsilon
\rangle = D(T) = \frac{\hbar\omega_a}{2} \coth \left(
\frac{\hbar\omega_a}{2k_B T} \right)$. We recall that
$\langle \epsilon \rangle = k_B T$ when the temperature is high
enough.

\section{Comparison with experimental data and conclusion}

In order to illustrate, in a quantitative manner, the great analogy
between the quantum tunneling description and our classical
stochastic escape rate calculation, the experimental results of Alberding et
al. \cite{alberding} will be used. Notice that
\begin{equation}\label{gma}
\frac{\gamma}{\omega_b} \ll
\frac{\gamma}{\omega_a} = \frac{e^2}{\hbar c} \frac{\hbar \omega_a}{m
c^2} \ll 1 ,
\end{equation}
so that the expression (\ref{lowk2}) can be written in the form
\begin{equation}\label{22}
\kappa(T) = \frac{\omega_a}{2 \pi} \exp \left(
-\frac{\Delta U}{\frac{\hbar\omega_a}{2}
\coth \left( \frac{\hbar\omega_a}{2k_B T} \right)} \right) .
\end{equation}

According to Alberding et al., the
beta-chain of hemoglobin ($\beta Hb$) is bounded to the carbon
monoxide $CO$ from which it can be separated with a LASER. The
rate of recombination can be obtained experimentally.
The fraction $N(t)$ of the molecules that have
not been recombined with $CO$ is measured as a function of time.
Then, the time $\tau$, necessary to reduce $N(t)$ to 75\% of its
original value, is determined. It is assumed that
this recombination is a passage through the potential barrier and
a good estimate of the escape rate is $\kappa = 1/ \tau$.

\begin{figure}\label{fig2}
\centerline{\includegraphics[width=100mm]{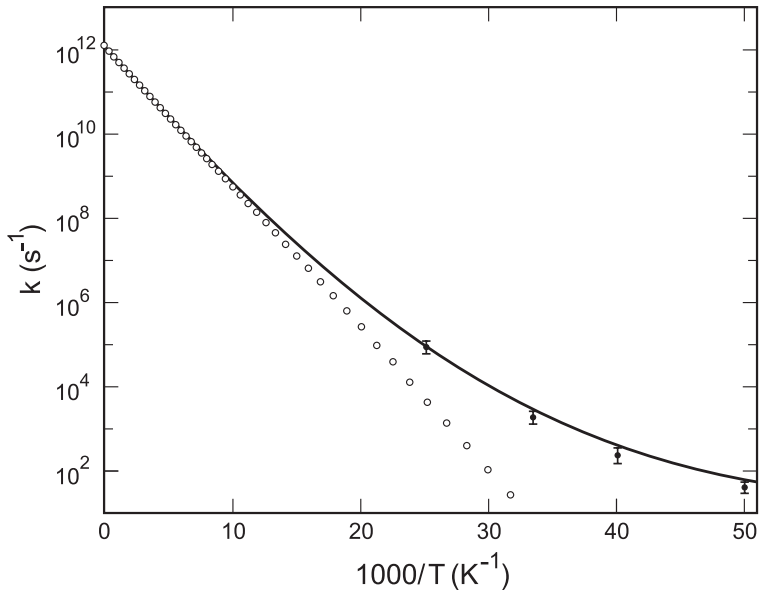}}
\caption{The solid line gives the high temperature behavior of our
theoretical prediction (formula (\ref{22})). The intermediate
temperature experimental data for the escape rate associated to
the CO migration to a separated $\beta$ chain of hemoglobin are
plotted (Alberding et al., 1976). The open circles corresponds to the
escape rate without the zero-point fluctuations, namely the Arrhenius
formula $\kappa = \frac{\omega_a}{2 \pi} \exp (-\Delta U/k_B T)$.}
\end{figure}

This experimental procedure can be repeated for different
temperatures $T$. The result for $\kappa(T)$ is indicated by the
experimental points (black dots) in the Figures 2 and 3, obtained by
N.R. Alberding and collaborators. We shall see that these experimental
data are very well described by the formula (\ref{22}).

We have adjusted  the values of $\omega_a$ and $\Delta U$ so that the
experimental data and the formula (\ref{22}) are in good
agreement. The values obtained are
\begin{eqnarray}
\frac{\hbar\omega_a}{2} = 2.53 \cdot 10^{-3} eV , \nonumber \\
\Delta U = 6.68 \cdot 10^{-2} eV .
\end{eqnarray}

Notice the impressive agreement between the classical theory with
zero-point radiation and the experimental data. This is more clearly
seen in the Figure 3. We conclude that the particle can escape from
the potential well at $T \rightarrow 0$, despite the fact that the
barrier height $\Delta U$ is much bigger than the particle mean
energy $\hbar\omega_a/2$ inside the well ($2 \Delta U / \hbar\omega_a
\approx 26$).

\begin{figure}\label{fig3}
\centerline{\includegraphics[width=100mm]{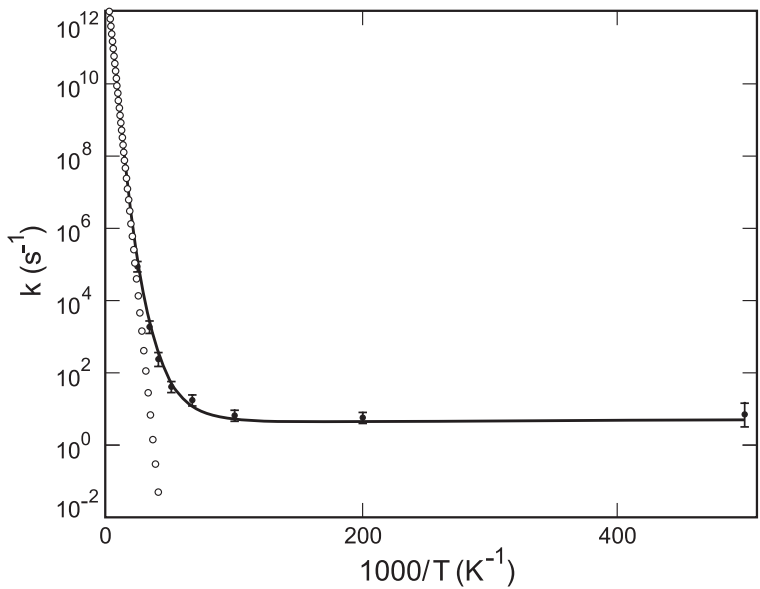}}
\caption{Experimental data for the escape rate associated to the
CO migration to a separated $\beta$ chain of hemoglobin (Alberding et
al., 1976). The solid line is our theoretical result (see
formula (\ref{22})). The open circles correspond to the escape rate
without the zero-point fluctuations (Arrhenius formula).}
\end{figure}

It is interesting to recall that Alberding et al. have obtained
$\Delta U$ using a conventional quantum mechanics calculation. They
have found a value for $\Delta U$ which is in semi-quantitative
agreement with the value obtained by us. However, the frequency
$\omega_a$ was not obtained by Alberding et al.. We want to stress
that the frequency $\omega_a$ gives relevant information about the
potential well (see (\ref{Ua})).

The classical stochastic interpretation of the zero temperature
escape rate is that the zero-point fluctuations provide enough
energy so that the particle can go over the potential barrier.
Since the particle is subjected to both the fluctuation and the
dissipation processes associated with the radiation bath, the energy
is not a constant of the motion. Therefore, particles that are
initially inside the potential well can escape and be detected at
points $x_c \gg x_b$ (see Figure 1), with a fluctuating energy
$\epsilon(t) < \Delta U$, contrary to the criticism of Baublitz
concerning the SED type calculation \cite{Baublitz}. Therefore, the
classical escape rate calculation presented in our paper gives results
{\em entirely analogous to the quantum tunneling description},
provided that the electromagnetic zero-point fluctuations are included
in the calculations.

\section*{Acknowledgements}
We thank the financial support from Funda\c{c}\~{a}o de Amparo \`{a}
Pesquisa do Estado de S\~{a}o Paulo (FAPESP) and Conselho Nacional de
Desenvolvimento Cient\'{i}fico e Tecnol\'{o}gico (CNPq-Brazil). We
also thank Prof. C. P. Malta for valuable comments.


\begin{thebibliography}{}


\bibitem{Kramers} H.A. Kramers, ``Brownian motion in a field of force
and the diffusion model of chemical reactions'', Physica 7 (1940) 284.

\bibitem{Hanggi} P. H\"{a}nggi, P. Talkner and M. Borkovec,
``Reaction-rate theory: fifty years after Kramers'', Rev. Mod. Phys. 62
(1990) 251.

\bibitem{Santos} T.W. Marshall and E. Santos, ``Semiclassical
treatment of macroscopic quantum relaxation'', Anales de Fisica 91
(1995) 49.

\bibitem{Marshall} T.W. Marshall, ``Random electrodynamics'',
Proc. Royal Soc. London 276A (1963) 475.

\bibitem{Boyer1} T.H. Boyer, ``Random electrodynamics: The theory of
classical electrodynamics with classical electromagnetic zero-point
radiation'', Phys. Rev. D 11 (1975) 790.

\bibitem{Boyer2} T.H. Boyer, ``A Brief Survey of Stochastic
Electrodynamics'' in: A.O. Barut, editor, Foundations of Radiation
Theory and Quantum Electrodynamics, pg. 49. Plenum Press, New York, 1980.

\bibitem{Pena2} L. de la Pe\~{n}a and A.M. Cetto, The Quantum Dice. An
Introduction to the Stochastic Electrodynamics. Kluwer Academic,
Dordrecht, 1996.

\bibitem{Milonni1} P.W. Milonni, ``Semiclassical and
quantum-electrodynamical approaches in nonrelativistic radiation
theory'', Phys. Rep. 25 (1976) 1.

\bibitem{jackson} J.D. Jackson, Classical Electrodynamics, 2nd
ed. Jonh Wiley \& Sons, New York, 1975. Chapters 9 and 17.

\bibitem{Chandrasekhar} S. Chandrasekhar, ``Stochastic Problems in
Physics and Astronomy'', Rev. Mod. Phys. 15 (1943) 1.

\bibitem{alberding} N. Alberding, R.H. Austin, K.W. Beeson, S.S. Chan,
L. Eisenstein, H. Frauenfelder and T.M. Nordlund, ``Tunneling in
ligand binding to heme proteins'', Science 192 (1976) 1002.

\bibitem{Baublitz} M. Baublitz Jr., ``Electron field-emission data,
quantum mechanics, and the classical stochastic theories'',
Phys. Rev. A 51 (1995) 1677.

\end{thebibliography}
\end{document}